\documentclass[11pt]{article}
\pdfoutput=1

\usepackage[margin=1in, paperwidth=8.5in, paperheight=11in]{geometry}

\usepackage{graphicx}
\usepackage{float}
\usepackage{subcaption}
\usepackage{fp}
\usepackage{pgfplots}
\pgfplotsset{compat=1.13}

\usepackage{tikz}

\usepackage{verbatim}

\usepackage[numbers]{natbib}
\usepackage{url}

\begin{document}

\title{ Improving the Accuracy of an Adiabatic Quantum Computer }
\author{John E. Dorband\\
	Department of Computer Science and Electrical Engineering\\
	University of Maryland, Baltimore County\\
	Maryland, USA\\
	\texttt{dorband@umbc.edu}}
\date{\today}
\maketitle

\begin{abstract}
The purpose of the D-Wave adiabatic quantum computer is to find a set of qubit values that minimize
its objective function.  
For various reasons, the set of qubit values returned by the D-Wave has errors.
This paper presents a method of improving the results returned by the D-Wave.
The method individually modifies the qubit values returned by the D-Wave to find a set of values which is a minimum 
of the objective function.
That set however is not necessarily guaranteed to be a global minimum. 
The method is simple and easily incorporated into any algorithm that has direct access to the sets of values 
returned by the D-Wave.  
Examples are also presented that demonstrate the merit of using such a sample improvement method.
\end{abstract}

\section{Introduction}\label{sec:intro}

The D-Wave\citep{Dwave13} is an adiabatic quantum computer\citep{Farhi00,Giuseppe08}.
The problem class that is addressed by the D-Wave is based on the Ising model objective function, F:
\begin{equation}\label{eq:obfunc}
F = {\sum\limits_i a_i q_i + \sum\limits_i \sum\limits_j b_{ij} q_i q_j}
\end{equation}
where $q_i\in\{-1,1\}$ are the qubit values returned by the D-Wave, and $a_i\in[-2,2]$ and $b_{ij}\in[-1,1]$ are 
the coefficients given to the D-Wave associated with the qubits and the qubit couplers respectively. 
A D-Wave 2x can have as many as 1152 qubit coefficients and 3360 coupler coefficients.
The C12 at NASA Ames has only 1097 qubit coefficients and 3060 coupler coefficients due primarily to
trapped magnetic flux.
Therefore to utilizing the D-Wave one must come up with a set of coefficients and send them to the D-Wave.
The D-Wave then returns at least one set of qubit values. 
This set is referred to here as a sample.
A request may be for thousands of samples. 
A typical request here results in 1000 samples.
The D-Wave's purpose is to return the set of qubit values which minimize F. 
There can only be one global minimum value, though there may be multiple samples with that global minimum.  
This global minimum value corresponds to the ground state of the D-Wave for the given set of coefficients.  
The D-Wave often returns a non-minimum energy state due to inherent quantum noise in the system, and the closeness 
of a large number of slightly higher energy 'active' states near the ground state.  
This leads to the question: if the D-Wave does not always return qubit values corresponding to the ground 
state what are the properties of the D-wave that can be depended upon to perform useful computations. 
Previously the author address this issue by characterizing the behavior of the D-Wave with various test 
cases\citep{Dorband16}.
The goal of this paper is to describe a method for correcting each sample returned by the D-Wave through post 
processing of each sample.  
Correction is defined here as determining a new set of qubit values which result 
in a minimum value of the objective function for the given set of coefficients ($a_i$s and $b_{ij}$s) 
based on the result returned by the D-Wave.
Once again this however does not guarantee that the correction will result in the global minimum value. 

\section{D-Wave Result Correction Method}\label{sec:method}

Each qubit $q_i$ has a domain of influence over the value of F.  
This influence consists of the qubit's coefficient $a_i$ and the coefficients  $b_{ij}$ of 
the couplers attached to it. Equation \ref{eq:influence} represents this influence.
\begin{equation}\label{eq:influence}
I_i = { a_i + \sum\limits_{(i,j)\in C} b_{ij} q_j}
\end{equation}
where $(i,j)$ is a coupler between qubit $q_i$ and $q_j$, and C is the set of couplers available on the D-Wave.  
The D-Wave is configured as a Chimera graph (see figure \ref{fig:chimera}) 
and some of those couples may not be functional due to the previously state trapped flux.

\begin{figure}
\centering
\begin{minipage}{0.5\textwidth}
	\centering
	\scalebox{1}{% A Chimera graph
% Author: John E. Dorband

% Chimera Graph Preamble (required
%\usepackage{tikz}

\begin{tikzpicture}[
  qubit/.style={circle, fill=black, thick, minimum size=2pt},
  scale=0.30,
  inner sep=0pt,
  ]

  \newcount\qu
  \newcount\x
  \newcount\y
  \newcount\xe
  \newcount\xf
  \newcount\ye
  \newcount\yf

  % Add east/west edges between cells
  \foreach \shd in {0,1,...,15} {
      \foreach \shr in {1,3,5} {
	  \draw [red] (\shr,\shd) .. controls (\shr + 0.5,\shd + 0.5) and (\shr + 1.5,\shd + 0.5) .. (\shr + 2,\shd);
      }
  }

  % Add north/south edges between cells
  \foreach \shd in {0,1,...,11} {
      \foreach \shr in {0,2,4,6} {
	  \draw [blue] (\shr,\shd) .. controls (\shr - .75,\shd + 1.0) and (\shr - .75,\shd + 2.5) .. (\shr,\shd + 4);
      }
  }

  \foreach \shd in {0,4,...,12} {
      \foreach \shr in {0,2,...,6} {
	  \foreach \xa in {0,...,1} {
	      \foreach \ya in {0,...,3} {

		  \x=\shr
		  \advance\x by \xa

		  \y=\shd
		  \advance\y by \ya

		  \qu=\x
		  \multiply\qu by 4
		  \advance\qu by \y

		  \node[qubit] (N-\the\qu) at (\x,\y) {};
	      }
	  }
	  \xe=\shr
	  \xf=\xe
	  \advance\xf by 1
	  \foreach \yg in {0,...,3} {
	      \foreach \yh in {0,...,3} {
		  \ye=\shd
		  \advance\ye by \yg
		  \yf=\shd
		  \advance\yf by \yh
		  \draw (\xe,\ye) -- (\xf,\yf);
	      }
	  }
      }
  }

\end{tikzpicture}}
	\caption[ISING DW CvsQ]{Chimera Graph(4x4 cells).}
	\label{fig:chimera}
\end{minipage}%
\end{figure}

Note that a set of $I_i$ is dependent on a specific sample returned from the D-Wave.
Since the goal is to minimize F if $I_i$ and $q_i$ have a differ sign for every qubit,
then F is a minimum, not necessarily the global minimum.
Therefore if any $I_i$ and $q_i$ have the same sign F is not a minimum and $q_i$ can be replaced by
$-q_i$ which will reduce the value of F by $2I_i$ for the modified sample. 
By recomputing $I_i$ for the new sample it can be determined if another $q_i$ may be modified to
reduce the value of F.
This can be repeated until the value of F can no longer be reduce (i.e. F is minimum).

\section{Correction Examples}\label{sec:examples}

D-Wave sample correction is simple and can be easily incorporated into any algorithm that
has direct access to samples returned by the D-Wave.
In the following sections the results from 3 algorithms are presented which utilizing the D-Wave 
with corrected samples as well as without.
The first example is a study of minimizing F for a set of random value qubit and coupler coefficients.
The second is that of training a chimera Boltzmann machine\citep{Dorband15} trained on 
hand written digits from the MNIST data set.
And finally the study on D-Wave characteristics\citep{Dorband16} is expanded from 
just comparing D-Wave results to results from a theoretically perfect D-Wave to including a D-Wave with
corrected results.

\subsection{Samples based on random coefficients}\label{sec:randomCoeff}

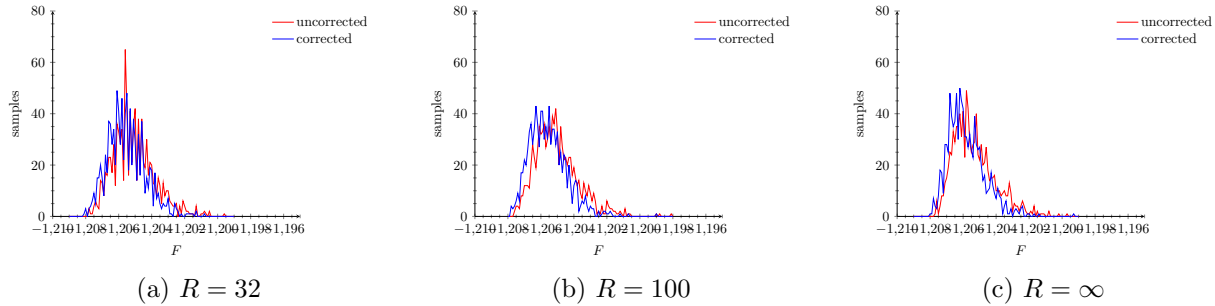
\begin{figure}
    \begin{subfigure}{0.32\textwidth}
	\scalebox{0.48}{
	    \begin{tikzpicture}
		  \pgfplotstableread{
"energy" "uncorrected" "corrected"
-1209 0 0
-1208.9 0 0
-1208.8 0 0
-1208.7 0 0
-1208.6 0 0
-1208.5 0 0
-1208.4 0 0
-1208.3 0 0
-1208.2 0 0
-1208.1 0 1
-1208 0 3
-1207.9 0 0
-1207.8 3 3
-1207.7 1 4
-1207.6 1 6
-1207.5 4 9
-1207.4 6 5
-1207.3 4 15
-1207.2 3 15
-1207.1 14 20
-1207 13 15
-1206.9 8 8
-1206.8 17 24
-1206.7 16 17
-1206.6 23 37
-1206.5 23 36
-1206.4 17 28
-1206.3 28 34
-1206.2 12 20
-1206.1 36 49
-1206 33 41
-1205.9 29 28
-1205.8 34 46
-1205.7 14 22
-1205.6 65 40
-1205.5 42 48
-1205.4 16 18
-1205.3 41 42
-1205.2 21 20
-1205.1 35 38
-1205 42 29
-1204.9 14 14
-1204.8 38 32
-1204.7 18 16
-1204.6 38 37
-1204.5 21 22
-1204.4 19 9
-1204.3 30 15
-1204.2 16 11
-1204.1 21 19
-1204 20 18
-1203.9 13 4
-1203.8 12 17
-1203.7 8 7
-1203.6 15 9
-1203.5 12 5
-1203.4 5 3
-1203.3 13 5
-1203.2 8 4
-1203.1 10 4
-1203 10 7
-1202.9 6 1
-1202.8 5 1
-1202.7 4 0
-1202.6 4 5
-1202.5 5 3
-1202.4 0 2
-1202.3 4 0
-1202.2 1 1
-1202.1 6 0
-1202 4 0
-1201.9 2 1
-1201.8 2 1
-1201.7 1 1
-1201.6 1 1
-1201.5 1 1
-1201.4 0 0
-1201.3 4 2
-1201.2 0 0
-1201.1 0 0
-1201 1 0
-1200.9 1 0
-1200.8 2 1
-1200.7 1 0
-1200.6 0 0
-1200.5 2 0
-1200.4 0 0
-1200.3 0 0
-1200.2 0 0
-1200.1 0 0
-1200 0 0
-1199.9 0 0
-1199.8 0 0
-1199.7 0 0
-1199.6 1 0
-1199.5 0 0
-1199.4 0 0
-1199.3 0 0
-1199.2 0 0
-1199.1 0 0
-1199 0 0
  }\plotdata;
		  \begin{axis} [
      axis lines=left,
      xlabel={$F$},
      ylabel={samples},
      ymin=0, ymax=80,
      xmin=-1210, xmax=-1195,
      minor x tick num=3,
      minor y tick num=3,
      legend cell align=left,
      legend style={ draw=none, at={(1.3,1.0)}, },
    ]
    \addplot[red]    table[ x index=0, y index=1, ]                {\plotdata};
    \addplot[blue]   table[ x index=0, y index=2, ]                {\plotdata};
    \legend{uncorrected,corrected}
  \end{axis}
	    \end{tikzpicture}
	}
	\caption{ $R=32$ } \label{fig:energyR32}
    \end{subfigure}
    \hspace*{\fill} % separation between the subfigures
    \begin{subfigure}{0.32\textwidth}
	\scalebox{0.48}{
	    \begin{tikzpicture}
		  \pgfplotstableread{
"energy" "uncorrected" "corrected"
-1208 0 0
-1207.9 0 0
-1207.8 0 4
-1207.7 0 3
-1207.6 2 4
-1207.5 4 6
-1207.4 4 9
-1207.3 3 6
-1207.2 8 17
-1207.1 8 17
-1207 12 22
-1206.9 12 20
-1206.8 12 26
-1206.7 11 33
-1206.6 20 36
-1206.5 28 28
-1206.4 23 34
-1206.3 19 43
-1206.2 26 36
-1206.1 35 27
-1206 32 41
-1205.9 33 41
-1205.8 36 30
-1205.7 27 35
-1205.6 30 30
-1205.5 35 43
-1205.4 29 28
-1205.3 39 34
-1205.2 36 34
-1205.1 42 28
-1205 33 33
-1204.9 20 20
-1204.8 35 25
-1204.7 27 17
-1204.6 22 24
-1204.5 23 22
-1204.4 20 11
-1204.3 23 20
-1204.2 23 13
-1204.1 16 5
-1204 19 12
-1203.9 15 14
-1203.8 12 13
-1203.7 14 2
-1203.6 9 4
-1203.5 7 6
-1203.4 13 5
-1203.3 10 8
-1203.2 8 4
-1203.1 12 2
-1203 9 4
-1202.9 6 3
-1202.8 9 3
-1202.7 6 0
-1202.6 1 0
-1202.5 2 2
-1202.4 1 0
-1202.3 4 1
-1202.2 1 2
-1202.1 1 2
-1202 6 1
-1201.9 4 1
-1201.8 3 2
-1201.7 3 1
-1201.6 2 0
-1201.5 2 0
-1201.4 1 0
-1201.3 1 0
-1201.2 2 1
-1201.1 1 0
-1201 0 0
-1200.9 3 0
-1200.8 0 0
-1200.7 1 1
-1200.6 1 0
-1200.5 0 0
-1200.4 0 0
-1200.3 0 0
-1200.2 0 0
-1200.1 0 0
-1200 0 0
-1199.9 0 0
-1199.8 0 0
-1199.7 0 0
-1199.6 0 0
-1199.5 0 0
-1199.4 0 0
-1199.3 0 0
-1199.2 0 0
-1199.1 0 0
-1199 1 1
-1198.9 0 0
-1198.8 1 0
-1198.7 0 0
-1198.6 0 0
-1198.5 0 0
-1198.4 0 0
-1198.3 0 0
-1198.2 0 0
-1198.1 1 0
-1198 0 0
  }\plotdata;
		  \begin{axis} [
      axis lines=left,
      xlabel={$F$},
      ylabel={samples},
      ymin=0, ymax=80,
      xmin=-1210, xmax=-1195,
      minor x tick num=3,
      minor y tick num=3,
      legend cell align=left,
      legend style={ draw=none, at={(1.3,1.0)}, },
    ]
    \addplot[red]    table[ x index=0, y index=1, ]                {\plotdata};
    \addplot[blue]   table[ x index=0, y index=2, ]                {\plotdata};
    \legend{uncorrected,corrected}
  \end{axis}
	    \end{tikzpicture}
	}
	\caption{ $R=100$ } \label{fig:energyR100}
    \end{subfigure}
    \hspace*{\fill} % separation between the subfigures
    \begin{subfigure}{0.32\textwidth}
	\scalebox{0.48}{
	    \begin{tikzpicture}
		  \pgfplotstableread{
"energy" "uncorrected" "corrected"
-1209 0 0
-1208.9 0 0
-1208.8 0 0
-1208.7 0 0
-1208.6 0 0
-1208.5 0 0
-1208.4 0 0
-1208.3 0 0
-1208.2 0 0
-1208.1 0 0
-1208 0 1
-1207.9 0 1
-1207.8 0 7
-1207.7 1 4
-1207.6 6 3
-1207.5 1 8
-1207.4 4 18
-1207.3 8 17
-1207.2 8 9
-1207.1 13 28
-1207 15 28
-1206.9 19 25
-1206.8 25 48
-1206.7 24 39
-1206.6 33 35
-1206.5 29 37
-1206.4 35 48
-1206.3 35 30
-1206.2 40 50
-1206.1 31 45
-1206 41 42
-1205.9 23 31
-1205.8 49 31
-1205.7 42 27
-1205.6 29 32
-1205.5 25 26
-1205.4 28 23
-1205.3 29 39
-1205.2 40 28
-1205.1 24 26
-1205 22 27
-1204.9 28 20
-1204.8 20 14
-1204.7 21 16
-1204.6 27 9
-1204.5 18 10
-1204.4 14 11
-1204.3 15 17
-1204.2 15 12
-1204.1 16 9
-1204 10 7
-1203.9 10 11
-1203.8 10 8
-1203.7 8 7
-1203.6 9 1
-1203.5 10 4
-1203.4 8 6
-1203.3 8 1
-1203.2 13 1
-1203.1 9 3
-1203 2 0
-1202.9 5 2
-1202.8 4 3
-1202.7 4 2
-1202.6 2 1
-1202.5 4 3
-1202.4 6 4
-1202.3 4 0
-1202.2 2 0
-1202.1 1 1
-1202 1 1
-1201.9 2 0
-1201.8 1 0
-1201.7 4 1
-1201.6 0 0
-1201.5 2 0
-1201.4 0 1
-1201.3 1 0
-1201.2 1 0
-1201.1 1 0
-1201 0 0
-1200.9 0 0
-1200.8 2 0
-1200.7 0 0
-1200.6 0 0
-1200.5 0 0
-1200.4 0 0
-1200.3 0 0
-1200.2 1 0
-1200.1 0 0
-1200 0 0
-1199.9 0 0
-1199.8 0 0
-1199.7 1 0
-1199.6 0 0
-1199.5 0 0
-1199.4 0 0
-1199.3 0 1
-1199.2 0 0
-1199.1 0 0
-1199 0 0
  }\plotdata;
		  \begin{axis} [
      axis lines=left,
      xlabel={$F$},
      ylabel={samples},
      ymin=0, ymax=80,
      xmin=-1210, xmax=-1195,
      minor x tick num=3,
      minor y tick num=3,
      legend cell align=left,
      legend style={ draw=none, at={(1.3,1.0)}, },
    ]
    \addplot[red]    table[ x index=0, y index=1, ]                {\plotdata};
    \addplot[blue]   table[ x index=0, y index=2, ]                {\plotdata};
    \legend{uncorrected,corrected}
  \end{axis}
	    \end{tikzpicture}
	}
	\caption{ $R=\infty$ } \label{fig:energyDB}
    \end{subfigure}
     \caption{ Histogram of values of $F$ from the D-Wave for a set of random coefficients. }
    \label{fig:energyC}
\end{figure}

In \citep{King17}, King, et al. defines ruggedness as follows:
\begin{equation}\label{eq:ruggedness}
 \left\lbrace \frac{f}{R} | f \in \{ -R, -R+1, \ldots ,R-1,R \} \right\rbrace
\end{equation}
where a real number, $f=[-1,1]$ is in essence a quantized real number of resolution $R$.
$R=1000$ is a relatively fine resolution, $R=10$ is a coarse resolution and $R=\infty$ is 
a non-quantized real number.
Three sets of coefficients were generated for the same set of random coefficients where 
$R=\infty$, $R=100$, and $R=32$. 
1000 samples were obtained from the D-Wave for each R values.
Figure \ref{fig:energyC} is 3 plots of the histograms of the values of $F$ for each the values of $R$
using the same set of random coefficients.
Each contains a histograms for the corrected and uncorrected values of $F$.
The corrected sample histogram is shifted, though slightly, to smaller values of $F$ as one might expect.  
All corrected sample values of $F$ were at least a local minimum nearby the corresponding D-Wave sample.
'Nearby' is referring to Hamming distance.

\subsection{Chimera Boltzmann machine}\label{sec:ChimeraBoltzM}

\begin{figure}[H]
    \begin{subfigure}{0.49\textwidth}
	\scalebox{0.75}{
	    \begin{tikzpicture}
		  \pgfplotstableread{
"energy" "uncorrected" "corrected"
-200 0 0
-200.1 0 0
-200.2 0 0
-200.3 0 0
-200.4 0 0
-200.5 0 0
-200.6 0 0
-200.7 0 0
-200.8 0 0
-200.9 0 0
-201 0 0
-201.1 0 0
-201.2 0 0
-201.3 0 0
-201.4 0 0
-201.5 0 0
-201.6 0 0
-201.7 1 0
-201.8 1 0
-201.9 0 0
-202 0 0
-202.1 1 0
-202.2 1 0
-202.3 0 0
-202.4 1 0
-202.5 0 0
-202.6 2 0
-202.7 4 0
-202.8 2 0
-202.9 1 0
-203 7 0
-203.1 6 0
-203.2 2 0
-203.3 6 0
-203.4 6 0
-203.5 7 0
-203.6 13 0
-203.7 14 0
-203.8 11 0
-203.9 17 0
-204 15 0
-204.1 15 0
-204.2 17 0
-204.3 26 0
-204.4 22 0
-204.5 23 0
-204.6 21 0
-204.7 20 0
-204.8 30 0
-204.9 26 0
-205 37 0
-205.1 29 0
-205.2 25 0
-205.3 41 0
-205.4 41 0
-205.5 40 0
-205.6 29 0
-205.7 38 0
-205.8 37 0
-205.9 36 0
-206 31 0
-206.1 32 0
-206.2 31 0
-206.3 35 0
-206.4 33 0
-206.5 38 0
-206.6 12 0
-206.7 15 0
-206.8 17 0
-206.9 13 0
-207 14 0
-207.1 13 0
-207.2 10 0
-207.3 9 0
-207.4 8 0
-207.5 8 0
-207.6 0 0
-207.7 5 0
-207.8 2 0
-207.9 2 0
-208 1 0
-208.1 0 0
-208.2 0 0
-208.3 0 0
-208.4 0 0
-208.5 0 0
-208.6 0 0
-208.7 0 0
-208.8 0 1000
-208.9 0 0
-209 0 0
-209.1 0 0
-209.2 0 0
-209.3 0 0
-209.4 0 0
-209.5 0 0
-209.6 0 0
-209.7 0 0
-209.8 0 0
-209.9 0 0
-210 0 0
  }\plotdata;
		  \begin{axis} [
      axis lines=left,
      xlabel={$F$},
      ylabel={samples},
      ymin=0, ymax=50,
      xmin=-210, xmax=-200,
      minor x tick num=3,
      minor y tick num=3,
      legend cell align=left,
      legend style={ draw=none, at={(1.3,1.0)}, },
    ]
    \addplot[red]    table[ x index=0, y index=1, ]                {\plotdata};
    \addplot[blue]   table[ x index=0, y index=2, ]                {\plotdata};
    \legend{uncorrected,corrected}
  \end{axis}
	    \end{tikzpicture}
	}
	\caption{ Single sample. } \label{fig:energyCBM}
    \end{subfigure}
    \hspace*{\fill} % separation between the subfigures
    \begin{subfigure}{0.49\textwidth}
	\scalebox{0.75}{
	    \begin{tikzpicture}
		  \pgfplotstableread{
"Epoch" "CoupledTrain" "CorrectedTrain" "CoupledTest" "CorrectedTest"
0 0.13 0.15 0.15 0.14
1 0.11 0.08 0.1 0.07
2 0.12 0.14 0.08 0.09
3 0.12 0.22 0.13 0.15
4 0.11 0.24 0.1 0.15
5 0.18 0.39 0.23 0.2
6 0.15 0.43 0.16 0.25
7 0.21 0.5 0.21 0.25
8 0.26 0.51 0.23 0.3
9 0.27 0.57 0.26 0.3
10 0.33 0.62 0.27 0.31
11 0.37 0.66 0.31 0.34
12 0.46 0.67 0.3 0.35
13 0.49 0.7 0.32 0.38
14 0.49 0.75 0.38 0.4
15 0.56 0.75 0.38 0.42
16 0.52 0.77 0.38 0.42
17 0.59 0.79 0.38 0.44
18 0.6 0.81 0.41 0.44
19 0.65 0.83 0.41 0.44
20 0.65 0.82 0.44 0.43
21 0.71 0.8 0.43 0.47
22 0.71 0.8 0.43 0.5
23 0.73 0.83 0.48 0.47
24 0.73 0.84 0.45 0.47
25 0.75 0.84 0.46 0.47
26 0.77 0.85 0.47 0.5
27 0.76 0.86 0.48 0.49
28 0.77 0.84 0.52 0.51
29 0.75 0.85 0.51 0.5
30 0.77 0.85 0.51 0.48
31 0.79 0.86 0.5 0.49
32 0.78 0.88 0.53 0.48
33 0.79 0.89 0.57 0.45
34 0.81 0.88 0.53 0.47
35 0.81 0.88 0.55 0.46
36 0.83 0.88 0.53 0.46
37 0.81 0.89 0.59 0.47
38 0.84 0.9 0.53 0.48
39 0.83 0.91 0.55 0.5
40 0.87 0.95 0.54 0.5
41 0.87 0.94 0.55 0.5
42 0.87 0.95 0.56 0.5
43 0.87 0.95 0.56 0.52
44 0.86 0.96 0.58 0.51
45 0.89 0.96 0.55 0.54
46 0.88 0.96 0.59 0.56
47 0.9 0.96 0.57 0.55
48 0.87 0.96 0.6 0.56
49 0.89 0.96 0.58 0.56
50 0.87 0.97 0.61 0.56
51 0.91 0.97 0.58 0.57
52 0.88 0.97 0.62 0.59
53 0.91 0.97 0.61 0.59
54 0.89 0.96 0.6 0.6
55 0.92 0.96 0.59 0.58
56 0.9 0.97 0.62 0.59
57 0.91 0.97 0.6 0.59
58 0.9 0.97 0.6 0.6
59 0.92 0.97 0.6 0.59
60 0.9 0.97 0.62 0.57
61 0.91 0.98 0.6 0.58
62 0.9 0.98 0.63 0.58
63 0.91 0.98 0.64 0.57
64 0.93 0.98 0.66 0.57
65 0.94 0.98 0.62 0.56
66 0.95 0.98 0.64 0.58
67 0.95 0.99 0.64 0.58
68 0.95 0.99 0.65 0.58
69 0.97 0.99 0.64 0.58
70 0.96 0.99 0.65 0.58
71 0.99 0.99 0.63 0.58
72 0.98 0.99 0.65 0.59
73 0.98 0.99 0.65 0.59
74 0.97 0.99 0.65 0.58
75 0.97 0.99 0.67 0.56
76 0.99 0.99 0.63 0.56
77 0.96 0.99 0.68 0.57
78 0.98 0.99 0.65 0.58
79 0.97 0.99 0.67 0.57
80 0.98 0.99 0.67 0.57
81 0.96 0.99 0.68 0.57
82 0.98 1 0.67 0.56
83 0.98 1 0.68 0.56
84 0.99 1 0.65 0.56
85 0.99 1 0.67 0.58
86 0.99 1 0.67 0.59
87 0.98 1 0.66 0.59
88 0.98 1 0.68 0.59
89 0.98 1 0.69 0.59
90 0.99 1 0.68 0.59
91 0.99 1 0.68 0.59
92 0.98 1 0.69 0.6
93 0.99 1 0.68 0.59
94 0.99 1 0.69 0.58
95 0.99 1 0.64 0.59
96 0.99 1 0.69 0.59
97 0.99 1 0.68 0.59
98 0.99 1 0.68 0.6
99 1 1 0.68 0.61
  }\plotdata;
		  \begin{axis} [
      axis lines=left,
      xlabel={Epochs},
      ylabel={Accuracy},
      ymin=0, ymax=1,
      xmin=0, xmax=100,
      minor x tick num=3,
      minor y tick num=3,
      legend cell align=left,
      legend style={ draw=none, at={(1.2,0.5)}, },
    ]
    \addplot[red]    table[ x index=0, y index=1, ]                {\plotdata};
    \addplot[orange] table[ x index=0, y index=2, ]                {\plotdata};
    \addplot[blue]   table[ x index=0, y index=3, ]                {\plotdata};
    \addplot[green]  table[ x index=0, y index=4, ]                {\plotdata};
    \legend{{Uncorrected Training},{Corrected Training},{Uncorrected Test},{Corrected Test}}
  \end{axis}
	    \end{tikzpicture}
	}
	\caption{ Training profile. } \label{fig:correctedCBM}
    \end{subfigure}
     \caption{ Result of Chimera Boltzmann machine while using uncorrected and corrected results form D-Wave }
    \label{fig:CBM}
\end{figure}
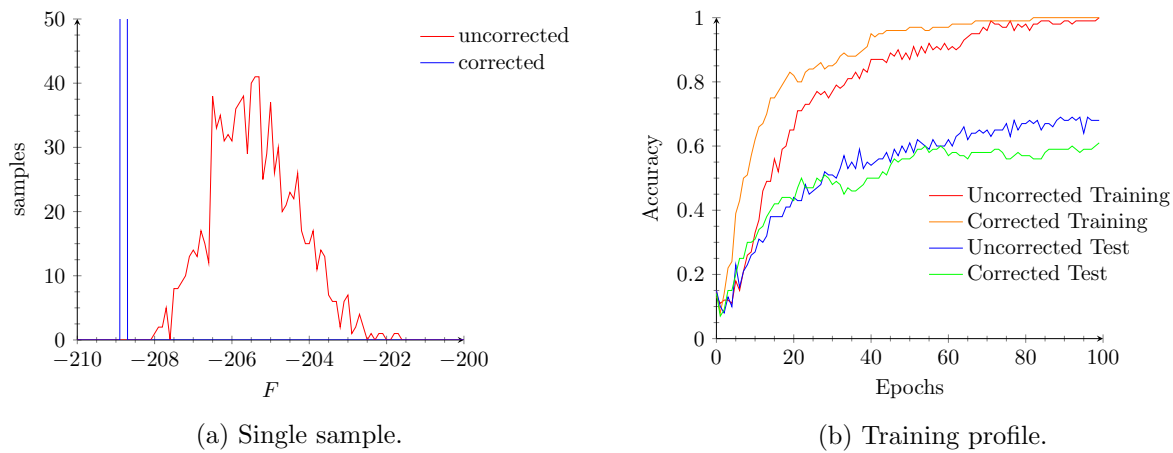

This example is performed on a Chimera Boltzmann Machine\citep{Dorband15} (CBM),
a 3 layer neural network, 2 visible and 1 hidden layer, base on the concepts of a Boltzmann machine.
Only the hidden layer is performed on the D-Wave.
It is being referred to as a chimera Boltzmann machine(CBM) because the hidden layer is connected like the
D-Wave (a chimera graph) instead of a restricted Boltzmann machine (RBM) which has no connections 
within the its hidden layers, and yet it is not a completely connected graph as a Boltzmann machine (BM)
would be.
This CBM is trained on 100 samples of hand written digits from the MNIST dataset for 200 epochs and tested on a different 
100 samples for the MNIST dataset.

Figure \ref{fig:energyCBM} is a histogram which shows the distribution of the values of $F$ for a typical
set of coefficients given to the D-Wave in the process of training the CBM.
The D-Wave returned 1000 samples for each set of coefficients.
Note that the distribution of uncorrected values form a bell-like shape curve and yet the corrected
value are all the same value (lower than the uncorrected values).
Though it is not proven, the correct values are arguably the global minimum.
In the process of training this CBM 40,000 requests were made to the D-Wave and in all cases the 1000 samples
for each request resulted in a single corrected minimum value, as in figure \ref{fig:energyCBM}.

Figure \ref{fig:correctedCBM} is the learning profile for the CBM for 100 training samples and 100 test samples
over 100 epochs.
Note that the CBM learned more rapidly using corrected samples than with uncorrected samples and yet the corrected
samples tended to lead to a slight over-fitting relative to the uncorrected samples.

\subsection{Virtual qubit characteristics}\label{sec:virtualQubit}

This example is an extension of the work done in reference \citep{Dorband16}.
For this example only results from an Ising model are given since the D-Wave is Ising model hardware.
For the Ising model the value of a qubit, $q_i$ can only have a value of -1 or 1.
A virtual qubit is a group of physical qubits that are treated and should act like a single physical qubit.
This means that if one qubit of the group has a value of 1 all the qubits of the group should have a 
value of 1 and if one qubit of the group does not have a value of 1, none of the qubits of the group 
should have a value of 1.
This is seldom the case.

To treat all the qubits symmetrically the same, all qubit coefficients will have the same value and 
all the couplers between qubits within the group will have the same coefficient values, 
yet the qubit coefficients and coupler coefficients may be different.

\begin{figure}[H]
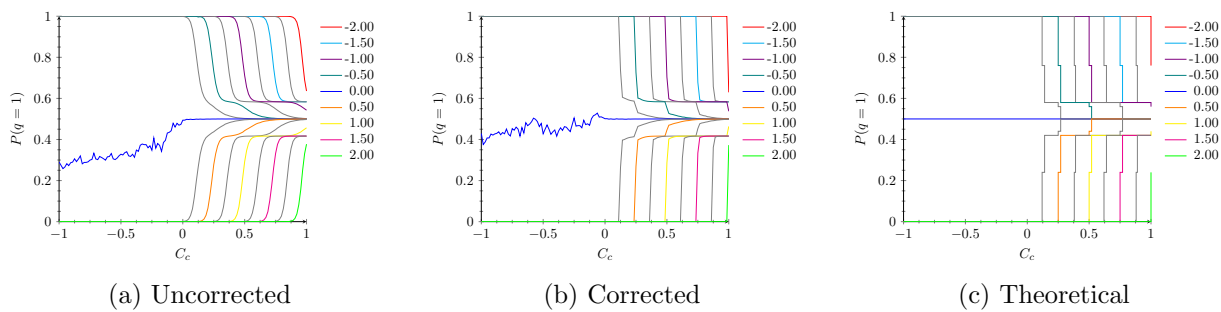

    \begin{subfigure}{0.32\textwidth}
	\scalebox{0.48}{
	    \begin{tikzpicture}
		\input{tab/chain_ISING_dwave_017_129}
		  \begin{axis} [
      axis lines=left,
      xlabel={$C_c$},
      ylabel={$P(q=1)$},
      ymin=0, ymax=1,
      xmin=-1, xmax=1,
      minor x tick num=3,
      minor y tick num=3,
      legend cell align=right,
      legend style={ draw=none, at={(1.3,1.0)}, },
    ]
    \addplot[red]    table[ x index=1, y index=2,  ]                {\plotdata};
    \addplot[gray]   table[ x index=1, y index=3,  forget plot, ]   {\plotdata};
    \addplot[cyan]   table[ x index=1, y index=4,  ]                {\plotdata};
    \addplot[gray]   table[ x index=1, y index=5,  forget plot, ]   {\plotdata};
    \addplot[violet] table[ x index=1, y index=6,  ]                {\plotdata};
    \addplot[gray]   table[ x index=1, y index=7,  forget plot, ]   {\plotdata};
    \addplot[teal]   table[ x index=1, y index=8,  ]                {\plotdata};
    \addplot[gray]   table[ x index=1, y index=9,  forget plot, ]   {\plotdata};
    \addplot[blue]   table[ x index=1, y index=10, ]                {\plotdata};
    \addplot[gray]   table[ x index=1, y index=11, forget plot, ]   {\plotdata};
    \addplot[orange] table[ x index=1, y index=12, ]                {\plotdata};
    \addplot[gray]   table[ x index=1, y index=13, forget plot, ]   {\plotdata};
    \addplot[yellow] table[ x index=1, y index=14, ]                {\plotdata};
    \addplot[gray]   table[ x index=1, y index=15, forget plot, ]   {\plotdata};
    \addplot[magenta]table[ x index=1, y index=16, ]                {\plotdata};
    \addplot[gray]   table[ x index=1, y index=17, forget plot, ]   {\plotdata};
    \addplot[green]  table[ x index=1, y index=18, ]                {\plotdata};
    \legend{-2.00,-1.50,-1.00,-0.50,0.00,0.50,1.00,1.50,2.00}
  \end{axis}
	    \end{tikzpicture}
	}
	\caption{ Uncorrected } \label{fig:IsingRawCc}
    \end{subfigure}
    \hspace*{\fill} % separation between the subfigures
    \begin{subfigure}{0.32\textwidth}
	\scalebox{0.48}{
	    \begin{tikzpicture}
		\input{tab/chain_ISING_correct_017_129}
		  \begin{axis} [
      axis lines=left,
      xlabel={$C_c$},
      ylabel={$P(q=1)$},
      ymin=0, ymax=1,
      xmin=-1, xmax=1,
      minor x tick num=3,
      minor y tick num=3,
      legend cell align=right,
      legend style={ draw=none, at={(1.3,1.0)}, },
    ]
    \addplot[red]    table[ x index=1, y index=2,  ]                {\plotdata};
    \addplot[gray]   table[ x index=1, y index=3,  forget plot, ]   {\plotdata};
    \addplot[cyan]   table[ x index=1, y index=4,  ]                {\plotdata};
    \addplot[gray]   table[ x index=1, y index=5,  forget plot, ]   {\plotdata};
    \addplot[violet] table[ x index=1, y index=6,  ]                {\plotdata};
    \addplot[gray]   table[ x index=1, y index=7,  forget plot, ]   {\plotdata};
    \addplot[teal]   table[ x index=1, y index=8,  ]                {\plotdata};
    \addplot[gray]   table[ x index=1, y index=9,  forget plot, ]   {\plotdata};
    \addplot[blue]   table[ x index=1, y index=10, ]                {\plotdata};
    \addplot[gray]   table[ x index=1, y index=11, forget plot, ]   {\plotdata};
    \addplot[orange] table[ x index=1, y index=12, ]                {\plotdata};
    \addplot[gray]   table[ x index=1, y index=13, forget plot, ]   {\plotdata};
    \addplot[yellow] table[ x index=1, y index=14, ]                {\plotdata};
    \addplot[gray]   table[ x index=1, y index=15, forget plot, ]   {\plotdata};
    \addplot[magenta]table[ x index=1, y index=16, ]                {\plotdata};
    \addplot[gray]   table[ x index=1, y index=17, forget plot, ]   {\plotdata};
    \addplot[green]  table[ x index=1, y index=18, ]                {\plotdata};
    \legend{-2.00,-1.50,-1.00,-0.50,0.00,0.50,1.00,1.50,2.00}
  \end{axis}
	    \end{tikzpicture}
	}
	\caption{ Corrected } \label{fig:IsingCorrCc}
    \end{subfigure}
    \hspace*{\fill} % separation between the subfigures
    \begin{subfigure}{0.32\textwidth}
	\scalebox{0.48}{
	    \begin{tikzpicture}
		\input{tab/chain_ISING_theoretic_017_129}
		  \begin{axis} [
      axis lines=left,
      xlabel={$C_c$},
      ylabel={$P(q=1)$},
      ymin=0, ymax=1,
      xmin=-1, xmax=1,
      minor x tick num=3,
      minor y tick num=3,
      legend cell align=right,
      legend style={ draw=none, at={(1.3,1.0)}, },
    ]
    \addplot[red]    table[ x index=1, y index=2,  ]                {\plotdata};
    \addplot[gray]   table[ x index=1, y index=3,  forget plot, ]   {\plotdata};
    \addplot[cyan]   table[ x index=1, y index=4,  ]                {\plotdata};
    \addplot[gray]   table[ x index=1, y index=5,  forget plot, ]   {\plotdata};
    \addplot[violet] table[ x index=1, y index=6,  ]                {\plotdata};
    \addplot[gray]   table[ x index=1, y index=7,  forget plot, ]   {\plotdata};
    \addplot[teal]   table[ x index=1, y index=8,  ]                {\plotdata};
    \addplot[gray]   table[ x index=1, y index=9,  forget plot, ]   {\plotdata};
    \addplot[blue]   table[ x index=1, y index=10, ]                {\plotdata};
    \addplot[gray]   table[ x index=1, y index=11, forget plot, ]   {\plotdata};
    \addplot[orange] table[ x index=1, y index=12, ]                {\plotdata};
    \addplot[gray]   table[ x index=1, y index=13, forget plot, ]   {\plotdata};
    \addplot[yellow] table[ x index=1, y index=14, ]                {\plotdata};
    \addplot[gray]   table[ x index=1, y index=15, forget plot, ]   {\plotdata};
    \addplot[magenta]table[ x index=1, y index=16, ]                {\plotdata};
    \addplot[gray]   table[ x index=1, y index=17, forget plot, ]   {\plotdata};
    \addplot[green]  table[ x index=1, y index=18, ]                {\plotdata};
    \legend{-2.00,-1.50,-1.00,-0.50,0.00,0.50,1.00,1.50,2.00}
  \end{axis}
	    \end{tikzpicture}
	}
	\caption{ Theoretical } \label{fig:IsingTheoCc}
    \end{subfigure}
    \caption{ Plots of the probability that a physical qubit within a 
      virtual qubit will be one, $P(q=1)$, vs. coupler coefficient value, 
      $C_c$, using the Ising model for 12 qubit chains on C12 at NASA Ames.  (17 different values of $C_q$ were plotted.) }
    \label{fig:IsingCc}
\end{figure}

\begin{figure}
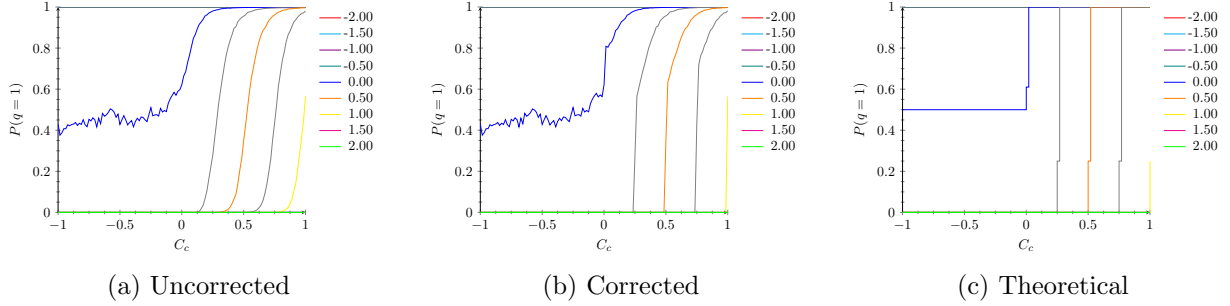

    \begin{subfigure}{0.32\textwidth}
	\scalebox{0.48}{
	    \begin{tikzpicture}
		\input{tab/chain_ISING_dwave_vote_017_129}
		  \begin{axis} [
      axis lines=left,
      xlabel={$C_c$},
      ylabel={$P(q=1)$},
      ymin=0, ymax=1,
      xmin=-1, xmax=1,
      minor x tick num=3,
      minor y tick num=3,
      legend cell align=right,
      legend style={ draw=none, at={(1.3,1.0)}, },
    ]
    \addplot[red]    table[ x index=1, y index=2,  ]                {\plotdata};
    \addplot[gray]   table[ x index=1, y index=3,  forget plot, ]   {\plotdata};
    \addplot[cyan]   table[ x index=1, y index=4,  ]                {\plotdata};
    \addplot[gray]   table[ x index=1, y index=5,  forget plot, ]   {\plotdata};
    \addplot[violet] table[ x index=1, y index=6,  ]                {\plotdata};
    \addplot[gray]   table[ x index=1, y index=7,  forget plot, ]   {\plotdata};
    \addplot[teal]   table[ x index=1, y index=8,  ]                {\plotdata};
    \addplot[gray]   table[ x index=1, y index=9,  forget plot, ]   {\plotdata};
    \addplot[blue]   table[ x index=1, y index=10, ]                {\plotdata};
    \addplot[gray]   table[ x index=1, y index=11, forget plot, ]   {\plotdata};
    \addplot[orange] table[ x index=1, y index=12, ]                {\plotdata};
    \addplot[gray]   table[ x index=1, y index=13, forget plot, ]   {\plotdata};
    \addplot[yellow] table[ x index=1, y index=14, ]                {\plotdata};
    \addplot[gray]   table[ x index=1, y index=15, forget plot, ]   {\plotdata};
    \addplot[magenta]table[ x index=1, y index=16, ]                {\plotdata};
    \addplot[gray]   table[ x index=1, y index=17, forget plot, ]   {\plotdata};
    \addplot[green]  table[ x index=1, y index=18, ]                {\plotdata};
    \legend{-2.00,-1.50,-1.00,-0.50,0.00,0.50,1.00,1.50,2.00}
  \end{axis}
	    \end{tikzpicture}
	}
	\caption{ Uncorrected } \label{fig:IsingRawCcV}
    \end{subfigure}
    \hspace*{\fill} % separation between the subfigures
    \begin{subfigure}{0.32\textwidth}
	\scalebox{0.48}{
	    \begin{tikzpicture}
		\input{tab/chain_ISING_correct_vote_017_129}
		  \begin{axis} [
      axis lines=left,
      xlabel={$C_c$},
      ylabel={$P(q=1)$},
      ymin=0, ymax=1,
      xmin=-1, xmax=1,
      minor x tick num=3,
      minor y tick num=3,
      legend cell align=right,
      legend style={ draw=none, at={(1.3,1.0)}, },
    ]
    \addplot[red]    table[ x index=1, y index=2,  ]                {\plotdata};
    \addplot[gray]   table[ x index=1, y index=3,  forget plot, ]   {\plotdata};
    \addplot[cyan]   table[ x index=1, y index=4,  ]                {\plotdata};
    \addplot[gray]   table[ x index=1, y index=5,  forget plot, ]   {\plotdata};
    \addplot[violet] table[ x index=1, y index=6,  ]                {\plotdata};
    \addplot[gray]   table[ x index=1, y index=7,  forget plot, ]   {\plotdata};
    \addplot[teal]   table[ x index=1, y index=8,  ]                {\plotdata};
    \addplot[gray]   table[ x index=1, y index=9,  forget plot, ]   {\plotdata};
    \addplot[blue]   table[ x index=1, y index=10, ]                {\plotdata};
    \addplot[gray]   table[ x index=1, y index=11, forget plot, ]   {\plotdata};
    \addplot[orange] table[ x index=1, y index=12, ]                {\plotdata};
    \addplot[gray]   table[ x index=1, y index=13, forget plot, ]   {\plotdata};
    \addplot[yellow] table[ x index=1, y index=14, ]                {\plotdata};
    \addplot[gray]   table[ x index=1, y index=15, forget plot, ]   {\plotdata};
    \addplot[magenta]table[ x index=1, y index=16, ]                {\plotdata};
    \addplot[gray]   table[ x index=1, y index=17, forget plot, ]   {\plotdata};
    \addplot[green]  table[ x index=1, y index=18, ]                {\plotdata};
    \legend{-2.00,-1.50,-1.00,-0.50,0.00,0.50,1.00,1.50,2.00}
  \end{axis}
	    \end{tikzpicture}
	}
	\caption{ Corrected } \label{fig:IsingCorrCcV}
    \end{subfigure}
    \hspace*{\fill} % separation between the subfigures
    \begin{subfigure}{0.32\textwidth}
	\scalebox{0.48}{
	    \begin{tikzpicture}
		\input{tab/chain_ISING_theoretic_vote_017_129}
		  \begin{axis} [
      axis lines=left,
      xlabel={$C_c$},
      ylabel={$P(q=1)$},
      ymin=0, ymax=1,
      xmin=-1, xmax=1,
      minor x tick num=3,
      minor y tick num=3,
      legend cell align=right,
      legend style={ draw=none, at={(1.3,1.0)}, },
    ]
    \addplot[red]    table[ x index=1, y index=2,  ]                {\plotdata};
    \addplot[gray]   table[ x index=1, y index=3,  forget plot, ]   {\plotdata};
    \addplot[cyan]   table[ x index=1, y index=4,  ]                {\plotdata};
    \addplot[gray]   table[ x index=1, y index=5,  forget plot, ]   {\plotdata};
    \addplot[violet] table[ x index=1, y index=6,  ]                {\plotdata};
    \addplot[gray]   table[ x index=1, y index=7,  forget plot, ]   {\plotdata};
    \addplot[teal]   table[ x index=1, y index=8,  ]                {\plotdata};
    \addplot[gray]   table[ x index=1, y index=9,  forget plot, ]   {\plotdata};
    \addplot[blue]   table[ x index=1, y index=10, ]                {\plotdata};
    \addplot[gray]   table[ x index=1, y index=11, forget plot, ]   {\plotdata};
    \addplot[orange] table[ x index=1, y index=12, ]                {\plotdata};
    \addplot[gray]   table[ x index=1, y index=13, forget plot, ]   {\plotdata};
    \addplot[yellow] table[ x index=1, y index=14, ]                {\plotdata};
    \addplot[gray]   table[ x index=1, y index=15, forget plot, ]   {\plotdata};
    \addplot[magenta]table[ x index=1, y index=16, ]                {\plotdata};
    \addplot[gray]   table[ x index=1, y index=17, forget plot, ]   {\plotdata};
    \addplot[green]  table[ x index=1, y index=18, ]                {\plotdata};
    \legend{-2.00,-1.50,-1.00,-0.50,0.00,0.50,1.00,1.50,2.00}
  \end{axis}
	    \end{tikzpicture}
	}
	\caption{ Theoretical } \label{fig:IsingTheoCcV}
    \end{subfigure}
    \caption{ Plots of the probability that a virtual qubit will be 1 (using voting over physical qubits), 
      $P(q=1)$, vs. coupler coefficient value, 
      $C_c$, using the Ising model for 12 qubit chains on C12 at NASA Ames.  (17 different values of $C_q$ were plotted.) }
    \label{fig:IsingCcV}
\end{figure}

\begin{figure}
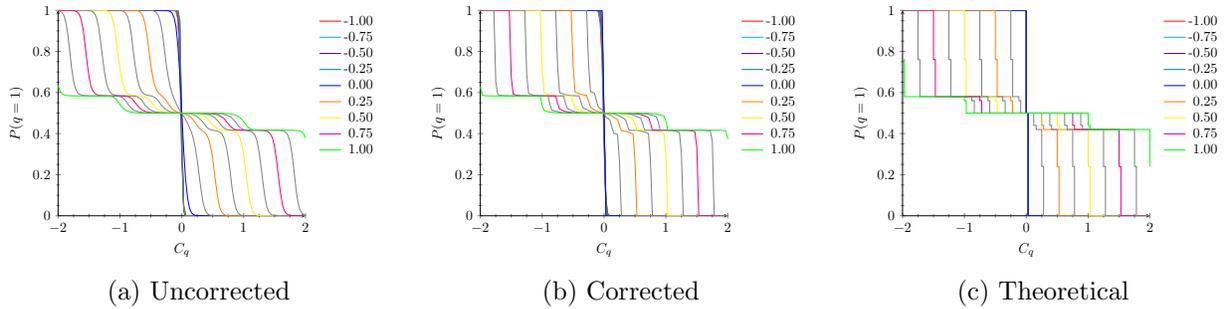

    \begin{subfigure}{0.32\textwidth}
	\scalebox{0.48}{
	    \begin{tikzpicture}
		\input{tab/chain_ISING_dwave_129_017}
		  \begin{axis} [
      axis lines=left,
      xlabel={$C_q$},
      ylabel={$P(q=1)$},
      ymin=0, ymax=1,
      xmin=-2, xmax=2,
      minor x tick num=3,
      minor y tick num=3,
      legend cell align=right,
      legend style={ draw=none, at={(1.3,1.0)}, },
    ]
    \addplot[red]    table[ x index=1, y index=2,  ]                {\plotdata};
    \addplot[gray]   table[ x index=1, y index=3,  forget plot, ]   {\plotdata};
    \addplot[cyan]   table[ x index=1, y index=4,  ]                {\plotdata};
    \addplot[gray]   table[ x index=1, y index=5,  forget plot, ]   {\plotdata};
    \addplot[violet] table[ x index=1, y index=6,  ]                {\plotdata};
    \addplot[gray]   table[ x index=1, y index=7,  forget plot, ]   {\plotdata};
    \addplot[teal]   table[ x index=1, y index=8,  ]                {\plotdata};
    \addplot[gray]   table[ x index=1, y index=9,  forget plot, ]   {\plotdata};
    \addplot[blue]   table[ x index=1, y index=10, ]                {\plotdata};
    \addplot[gray]   table[ x index=1, y index=11, forget plot, ]   {\plotdata};
    \addplot[orange] table[ x index=1, y index=12, ]                {\plotdata};
    \addplot[gray]   table[ x index=1, y index=13, forget plot, ]   {\plotdata};
    \addplot[yellow] table[ x index=1, y index=14, ]                {\plotdata};
    \addplot[gray]   table[ x index=1, y index=15, forget plot, ]   {\plotdata};
    \addplot[magenta]table[ x index=1, y index=16, ]                {\plotdata};
    \addplot[gray]   table[ x index=1, y index=17, forget plot, ]   {\plotdata};
    \addplot[green]  table[ x index=1, y index=18, ]                {\plotdata};
    \legend{-1.00,-0.75,-0.50,-0.25,0.00,0.25,0.50,0.75,1.00}
  \end{axis}
	    \end{tikzpicture}
	}
	\caption{ Uncorrected } \label{fig:IsingRawCq}
    \end{subfigure}
    \hspace*{\fill} % separation between the subfigures
    \begin{subfigure}{0.32\textwidth}
	\scalebox{0.48}{
	    \begin{tikzpicture}
		\input{tab/chain_ISING_correct_129_017}
		  \begin{axis} [
      axis lines=left,
      xlabel={$C_q$},
      ylabel={$P(q=1)$},
      ymin=0, ymax=1,
      xmin=-2, xmax=2,
      minor x tick num=3,
      minor y tick num=3,
      legend cell align=right,
      legend style={ draw=none, at={(1.3,1.0)}, },
    ]
    \addplot[red]    table[ x index=1, y index=2,  ]                {\plotdata};
    \addplot[gray]   table[ x index=1, y index=3,  forget plot, ]   {\plotdata};
    \addplot[cyan]   table[ x index=1, y index=4,  ]                {\plotdata};
    \addplot[gray]   table[ x index=1, y index=5,  forget plot, ]   {\plotdata};
    \addplot[violet] table[ x index=1, y index=6,  ]                {\plotdata};
    \addplot[gray]   table[ x index=1, y index=7,  forget plot, ]   {\plotdata};
    \addplot[teal]   table[ x index=1, y index=8,  ]                {\plotdata};
    \addplot[gray]   table[ x index=1, y index=9,  forget plot, ]   {\plotdata};
    \addplot[blue]   table[ x index=1, y index=10, ]                {\plotdata};
    \addplot[gray]   table[ x index=1, y index=11, forget plot, ]   {\plotdata};
    \addplot[orange] table[ x index=1, y index=12, ]                {\plotdata};
    \addplot[gray]   table[ x index=1, y index=13, forget plot, ]   {\plotdata};
    \addplot[yellow] table[ x index=1, y index=14, ]                {\plotdata};
    \addplot[gray]   table[ x index=1, y index=15, forget plot, ]   {\plotdata};
    \addplot[magenta]table[ x index=1, y index=16, ]                {\plotdata};
    \addplot[gray]   table[ x index=1, y index=17, forget plot, ]   {\plotdata};
    \addplot[green]  table[ x index=1, y index=18, ]                {\plotdata};
    \legend{-1.00,-0.75,-0.50,-0.25,0.00,0.25,0.50,0.75,1.00}
  \end{axis}
	    \end{tikzpicture}
	}
	\caption{ Corrected } \label{fig:IsingCorrCq}
    \end{subfigure}
    \hspace*{\fill} % separation between the subfigures
    \begin{subfigure}{0.32\textwidth}
	\scalebox{0.48}{
	    \begin{tikzpicture}
		\input{tab/chain_ISING_theoretic_129_017}
		  \begin{axis} [
      axis lines=left,
      xlabel={$C_q$},
      ylabel={$P(q=1)$},
      ymin=0, ymax=1,
      xmin=-2, xmax=2,
      minor x tick num=3,
      minor y tick num=3,
      legend cell align=right,
      legend style={ draw=none, at={(1.3,1.0)}, },
    ]
    \addplot[red]    table[ x index=1, y index=2,  ]                {\plotdata};
    \addplot[gray]   table[ x index=1, y index=3,  forget plot, ]   {\plotdata};
    \addplot[cyan]   table[ x index=1, y index=4,  ]                {\plotdata};
    \addplot[gray]   table[ x index=1, y index=5,  forget plot, ]   {\plotdata};
    \addplot[violet] table[ x index=1, y index=6,  ]                {\plotdata};
    \addplot[gray]   table[ x index=1, y index=7,  forget plot, ]   {\plotdata};
    \addplot[teal]   table[ x index=1, y index=8,  ]                {\plotdata};
    \addplot[gray]   table[ x index=1, y index=9,  forget plot, ]   {\plotdata};
    \addplot[blue]   table[ x index=1, y index=10, ]                {\plotdata};
    \addplot[gray]   table[ x index=1, y index=11, forget plot, ]   {\plotdata};
    \addplot[orange] table[ x index=1, y index=12, ]                {\plotdata};
    \addplot[gray]   table[ x index=1, y index=13, forget plot, ]   {\plotdata};
    \addplot[yellow] table[ x index=1, y index=14, ]                {\plotdata};
    \addplot[gray]   table[ x index=1, y index=15, forget plot, ]   {\plotdata};
    \addplot[magenta]table[ x index=1, y index=16, ]                {\plotdata};
    \addplot[gray]   table[ x index=1, y index=17, forget plot, ]   {\plotdata};
    \addplot[green]  table[ x index=1, y index=18, ]                {\plotdata};
    \legend{-1.00,-0.75,-0.50,-0.25,0.00,0.25,0.50,0.75,1.00}
  \end{axis}
	    \end{tikzpicture}
	}
	\caption{ Theoretical } \label{fig:IsingTheoQq}
    \end{subfigure}
    \caption{ Plots of the probability that a physical qubit within a 
      virtual qubit will be 1, $P(q=1)$, vs. qubit coefficient value, 
      $C_q$, using the Ising model for 12 qubit chains on C12 at NASA Ames.  (17 different values of $C_c$ were plotted.) }
    \label{fig:IsingCq}
\end{figure}

\begin{figure}
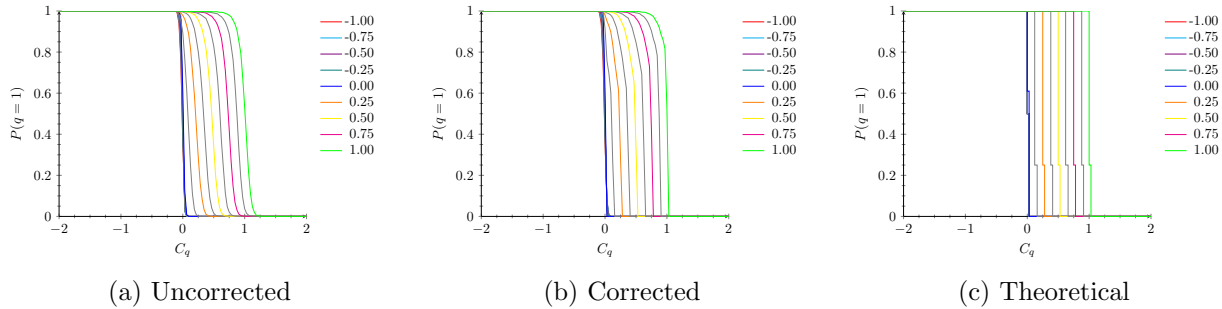

    \begin{subfigure}{0.32\textwidth}
	\scalebox{0.48}{
	    \begin{tikzpicture}
		\input{tab/chain_ISING_dwave_vote_129_017}
		  \begin{axis} [
      axis lines=left,
      xlabel={$C_q$},
      ylabel={$P(q=1)$},
      ymin=0, ymax=1,
      xmin=-2, xmax=2,
      minor x tick num=3,
      minor y tick num=3,
      legend cell align=right,
      legend style={ draw=none, at={(1.3,1.0)}, },
    ]
    \addplot[red]    table[ x index=1, y index=2,  ]                {\plotdata};
    \addplot[gray]   table[ x index=1, y index=3,  forget plot, ]   {\plotdata};
    \addplot[cyan]   table[ x index=1, y index=4,  ]                {\plotdata};
    \addplot[gray]   table[ x index=1, y index=5,  forget plot, ]   {\plotdata};
    \addplot[violet] table[ x index=1, y index=6,  ]                {\plotdata};
    \addplot[gray]   table[ x index=1, y index=7,  forget plot, ]   {\plotdata};
    \addplot[teal]   table[ x index=1, y index=8,  ]                {\plotdata};
    \addplot[gray]   table[ x index=1, y index=9,  forget plot, ]   {\plotdata};
    \addplot[blue]   table[ x index=1, y index=10, ]                {\plotdata};
    \addplot[gray]   table[ x index=1, y index=11, forget plot, ]   {\plotdata};
    \addplot[orange] table[ x index=1, y index=12, ]                {\plotdata};
    \addplot[gray]   table[ x index=1, y index=13, forget plot, ]   {\plotdata};
    \addplot[yellow] table[ x index=1, y index=14, ]                {\plotdata};
    \addplot[gray]   table[ x index=1, y index=15, forget plot, ]   {\plotdata};
    \addplot[magenta]table[ x index=1, y index=16, ]                {\plotdata};
    \addplot[gray]   table[ x index=1, y index=17, forget plot, ]   {\plotdata};
    \addplot[green]  table[ x index=1, y index=18, ]                {\plotdata};
    \legend{-1.00,-0.75,-0.50,-0.25,0.00,0.25,0.50,0.75,1.00}
  \end{axis}
	    \end{tikzpicture}
	}
	\caption{ Uncorrected } \label{fig:IsingRawCqV}
    \end{subfigure}
    \hspace*{\fill} % separation between the subfigures
    \begin{subfigure}{0.32\textwidth}
	\scalebox{0.48}{
	    \begin{tikzpicture}
		\input{tab/chain_ISING_correct_vote_129_017}
		  \begin{axis} [
      axis lines=left,
      xlabel={$C_q$},
      ylabel={$P(q=1)$},
      ymin=0, ymax=1,
      xmin=-2, xmax=2,
      minor x tick num=3,
      minor y tick num=3,
      legend cell align=right,
      legend style={ draw=none, at={(1.3,1.0)}, },
    ]
    \addplot[red]    table[ x index=1, y index=2,  ]                {\plotdata};
    \addplot[gray]   table[ x index=1, y index=3,  forget plot, ]   {\plotdata};
    \addplot[cyan]   table[ x index=1, y index=4,  ]                {\plotdata};
    \addplot[gray]   table[ x index=1, y index=5,  forget plot, ]   {\plotdata};
    \addplot[violet] table[ x index=1, y index=6,  ]                {\plotdata};
    \addplot[gray]   table[ x index=1, y index=7,  forget plot, ]   {\plotdata};
    \addplot[teal]   table[ x index=1, y index=8,  ]                {\plotdata};
    \addplot[gray]   table[ x index=1, y index=9,  forget plot, ]   {\plotdata};
    \addplot[blue]   table[ x index=1, y index=10, ]                {\plotdata};
    \addplot[gray]   table[ x index=1, y index=11, forget plot, ]   {\plotdata};
    \addplot[orange] table[ x index=1, y index=12, ]                {\plotdata};
    \addplot[gray]   table[ x index=1, y index=13, forget plot, ]   {\plotdata};
    \addplot[yellow] table[ x index=1, y index=14, ]                {\plotdata};
    \addplot[gray]   table[ x index=1, y index=15, forget plot, ]   {\plotdata};
    \addplot[magenta]table[ x index=1, y index=16, ]                {\plotdata};
    \addplot[gray]   table[ x index=1, y index=17, forget plot, ]   {\plotdata};
    \addplot[green]  table[ x index=1, y index=18, ]                {\plotdata};
    \legend{-1.00,-0.75,-0.50,-0.25,0.00,0.25,0.50,0.75,1.00}
  \end{axis}
	    \end{tikzpicture}
	}
	\caption{ Corrected } \label{fig:IsingCorrCqV}
    \end{subfigure}
    \hspace*{\fill} % separation between the subfigures
    \begin{subfigure}{0.32\textwidth}
	\scalebox{0.48}{
	    \begin{tikzpicture}
		\input{tab/chain_ISING_theoretic_vote_129_017}
		  \begin{axis} [
      axis lines=left,
      xlabel={$C_q$},
      ylabel={$P(q=1)$},
      ymin=0, ymax=1,
      xmin=-2, xmax=2,
      minor x tick num=3,
      minor y tick num=3,
      legend cell align=right,
      legend style={ draw=none, at={(1.3,1.0)}, },
    ]
    \addplot[red]    table[ x index=1, y index=2,  ]                {\plotdata};
    \addplot[gray]   table[ x index=1, y index=3,  forget plot, ]   {\plotdata};
    \addplot[cyan]   table[ x index=1, y index=4,  ]                {\plotdata};
    \addplot[gray]   table[ x index=1, y index=5,  forget plot, ]   {\plotdata};
    \addplot[violet] table[ x index=1, y index=6,  ]                {\plotdata};
    \addplot[gray]   table[ x index=1, y index=7,  forget plot, ]   {\plotdata};
    \addplot[teal]   table[ x index=1, y index=8,  ]                {\plotdata};
    \addplot[gray]   table[ x index=1, y index=9,  forget plot, ]   {\plotdata};
    \addplot[blue]   table[ x index=1, y index=10, ]                {\plotdata};
    \addplot[gray]   table[ x index=1, y index=11, forget plot, ]   {\plotdata};
    \addplot[orange] table[ x index=1, y index=12, ]                {\plotdata};
    \addplot[gray]   table[ x index=1, y index=13, forget plot, ]   {\plotdata};
    \addplot[yellow] table[ x index=1, y index=14, ]                {\plotdata};
    \addplot[gray]   table[ x index=1, y index=15, forget plot, ]   {\plotdata};
    \addplot[magenta]table[ x index=1, y index=16, ]                {\plotdata};
    \addplot[gray]   table[ x index=1, y index=17, forget plot, ]   {\plotdata};
    \addplot[green]  table[ x index=1, y index=18, ]                {\plotdata};
    \legend{-1.00,-0.75,-0.50,-0.25,0.00,0.25,0.50,0.75,1.00}
  \end{axis}
	    \end{tikzpicture}
	}
	\caption{ Theoretical } \label{fig:IsingTheoQqV}
    \end{subfigure}
    \caption{ Plots of the probability that a virtual qubit will be 1 (using voting over physical qubits), 
      $P(q=1)$, vs. qubit coefficient value, $C_q$, 
      using the Ising model for 12 qubit chains on C12 at NASA Ames.  (17 different values of $C_c$ were plotted.) }
    \label{fig:IsingCqV}
\end{figure}

Figures \ref{fig:IsingCc}-\ref{fig:IsingCqV} are families of curves where $C_c=[-1,1]$ and $C_q=[-2,2]$ 
are plotted against each other. Each triplet of plots represent a) uncorrected D-Wave samples, 
b) corrected D-Wave samples, and s) samples of a theoretically errorless D-Wave.
Figure \ref{fig:IsingCc} plots the probability that a qubit within a virtual qubit 
will have a value of 1 versus a coupler coefficient value, $C_c$, where each line is a different value 
of qubit coefficient, $C_q$.
Figure \ref{fig:IsingCq} plots the probability that a qubit within a virtual qubit 
will have a value of 1 versus $C_q$, where each line is a different value of $C_c$.

Voting can also used as a metric to reduce error.
Voting consists of counting the number of physical qubits in the virtual qubit that have a value of 1
and if that number is larger than (or equal) to half the number of physical qubits in the virtual qubit 
the value of the virtual qubit is consider to be 1.
Figures \ref{fig:IsingCcV} and \ref{fig:IsingCqV} are the same as \ref{fig:IsingCc} and \ref{fig:IsingCq} 
except that rather than plotting the probability of a qubit within a virtual qubit having a value of 1,
the probability of a virtual qubit having a value of 1 by voting is plotted.

Finally it should be noted that in all cases the plots using correct samples are much closer 
in appearance to the theoretically errorless D-Wave results than are the uncorrected D-Wave results.

\section{Conclusion}\label{sec:conclusion}

A method has been presented that utilizes the D-Wave, and is guaranteed, to find a minimum nearby any sample 
returned by the D-Wave,
However, the minimum is not guaranteed to be the global minimum.
It is simple and is easily incorporated into any algorithm that has access to the raw samples returned by
the D-Wave. 
It could be execution time optimized but this is not address here.
Three examples of the effects of using corrected D-wave samples have been presented and it can be easily seen
that the results have been improved, more in some cases than in others.

\section*{Acknowledgement}

The author would like to thank Michael Little and Marjorie Cole of the NASA Advanced Information Systems 
Technology Office for their continued support for this research effort under grant NNX15AK58G and to the 
NASA Ames Research Center for providing access to the D-Wave quantum annealing computer. 
In addition, the author thanks the NSF funded Center for Hybrid Multicore Productivity Research and 
D-Wave Systems for their support and access to their computational resources.  

\bibliography{QubitCorrection}{}

\end{document}